\documentclass[iop]{emulateapj}
\usepackage{longtable,natbib}
\usepackage[bookmarks=false]{hyperref}

\def\arcsec{{\mbox{$^{\prime \prime}$}}}

\def\degree{{\mbox{$^{\circ}$}}}

\begin{document}
\title{Understanding Dual AGN Activation in the Nearby Universe}

\author{Michael Koss\altaffilmark{1,2}, Richard Mushotzky\altaffilmark{2}, Ezequiel Treister\altaffilmark{3},  Sylvain Veilleux\altaffilmark{2}, Ranjan Vasudevan\altaffilmark{2}, and Margaret Trippe\altaffilmark{2}}
\email{koss@ifa.hawaii.edu}
\altaffiltext{1}{Institute for Astronomy, University of Hawaii, Honolulu, HI, USA}
\altaffiltext{2}{Astronomy Department, University of Maryland, College Park, MD, USA}
\altaffiltext{3}{Departamento de Astronom\'{\i}a, Universidad de Concepci\'{o}n, Concepci\'{o}n, Chile}

\begin{abstract}
We study the fraction of dual AGN in a sample of 167 nearby (z$<$0.05), moderate luminosity, ultra hard X-ray selected AGN from the all-sky $Swift$ BAT survey.  Combining new $Chandra$ and Gemini observations together with optical and X-ray observations, we find that the dual AGN frequency at scales $<$100 kpc is $\sim$10\% (16/167).  Of the 16 dual AGN, only 3 (19\%) were detected using X-ray spectroscopy and were not detected using emission line diagnostics.   Close dual AGN ($<$30 kpc) tend to be more common among the most X-ray luminous systems.  In dual AGN, the X-ray luminosity of both AGN increases strongly with decreasing galaxy separation, suggesting that the merging event is key in powering both AGN.  50\% of the AGN with a very close companion ($<$15 kpc), are dual AGN.  We also find that dual AGN are more likely to occur in major mergers and tend to avoid absorption line galaxies with elliptical morphologies.  Finally, we find SDSS Seyferts are much less likely than BAT AGN (0.25\% vs.~7.8\%) to be found in dual AGN at scales $<$30 kpc because of a smaller number of companions galaxies, fiber collision limits, a tendency for AGN at small separations to be detected only in X-rays, and a higher fraction of dual AGN companions with increasing AGN luminosity.  

\end{abstract}

\keywords{galaxies: active}

\section{Introduction}
	 
	The detection and frequency of dual active galactic nuclei (AGN) on kpc scales is an important test of the merger-driven AGN model.  Understanding the types of galaxies and specific merger stages where dual AGN occur provides important clues about the peak black hole growth during the merging process.  Over the last decade, several nearby dual AGN on kpc scales have been found serendipitously in interacting galaxies \citep{Komossa:2003p11219,Koss:2011p12483}.  In addition, higher redshift dual AGN systems have been discovered using high resolution integral field spectroscopy or radio observations \citep{Fu:2011p14318}.  Finding these duals is an important step, but a systematic study of dual AGN offers the best chance of understanding their frequency and the types of galaxies and interactions involved in their triggering.


	 
	 Early studies of the dual AGN frequency at higher redshift  (z$>$0.15) focused on quasar pairs but suffered from severe observational difficulties detecting close pairs.  \citet{Djorgovski:1991p15029} found a 2x quasar overabundance on $<$100 kpc scales based on normal galaxy clustering.  Other early observations found a small fraction (0.1\%) of quasar pairs with typical separations of 50 to 100 kpc suggesting that at least one AGN shuts off on scales $<$30 kpc \citep{Mortlock:1999p13970}.  Recent studies using the SDSS suggested dual quasars peak at small $<$50 kpc scales with a small excess extending to 100s of kpc \citep{Hennawi:2006p15127,Foreman:2009p14137}.  Other recent studies found the opposite result with only a small excess at $<$$35$ kpc \citep{Myers:2007p15297}.     However, on close scales $<$30 kpc, these observations are complicated by gravitational lenses \citep{Narayan:1988p14509,Kochanek:1999p15092} as well as an inability to easily resolve each AGN in optical and X-ray observations.  


	 

	Nearby galaxies are better suited for dual AGN studies since both the nuclear engine and the host galaxies can be resolved and studied down to separations of a few kpc.   A large study of optical AGN pairs at z$<$0.16 using SDSS spectroscopy found that their fraction was $\approx$1.5\% within 30 kpc \citep{Liu:2011p14491} while their cumulative fraction increases linearly with separation.  However, because of fiber collision limits this technique suffers from incompleteness at close scales.  Furthermore, optical surveys are incomplete since X-ray and IR selected AGN are not always detected optically \citep{Hickox:2009p15012,Koss:2011p12483}.  Therefore, a study of nearby AGN using emission line diagnostics and X-ray observations provides the best chance of accurately measuring the dual AGN frequency.    
	
	

	In this letter, we determine the incidence of dual AGN in the local Universe by searching for multiple active nuclei from the \citet{Koss:2011p14514} sample of ultra hard X-ray selected AGN.   This sample is ideal since archival X-ray data from $SWIFT$ XRT, $XMM$, or $Chandra$ exist for the entire set of objects.  The ultra hard X-rays also offer a reliable tracer of the bolometric luminosity of AGN since it is nearly unaffected by host galaxy contamination or obscuration, which strongly affects other selection techniques such as optical emission lines, infrared spectral analysis, or optical colors.
	


\section{Data and Analysis}
We adopt $\Omega_m$= 0.3, $\Omega_\Lambda$= 0.7, and $H_0$ = 70 km s$^{-1}$ Mpc$^{-1}$ to determine distances. 

\subsection{Samples}
	The BAT AGN sample consists of 167 nearby ($z$$<$$0.05$) ultra hard X-ray BAT-detected AGN from \citet{Koss:2011p14514} catalog in the Northern sky ($>$-25$\degree$), with low Galactic extinction ($E(B-V)$$<$0.5).  For each BAT AGN, we search the SDSS,  6DF, 2DF, and NED for apparent companions within 100 kpc with small radial velocity differences ($<$300 km s$^{-1}$). For comparison, we use a sample of emission line selected type 2 Seyferts in the SDSS taken from the Garching catalog, which we refer to as {\it SDSS Seyferts}. We use the emission line diagnostics of \citet{Veilleux:1987p1782}, revised by \citet{Kewley:2006p1554}.  We require each optical AGN to be classified as Seyfert using the [O III] $\lambda$5007/H$\beta$ vs.~[N II] $\lambda$6583/H$\alpha$, [S II] $\lambda\lambda$6717, 6731/H$\alpha$, and [O I] $\lambda$6300/H$\alpha$ diagnostics, as well as having a S/N$>$10 in all lines.  We restricted our SDSS Seyferts to $z$$<$0.07 totaling 1988 Seyferts.  

\subsection{X-ray Data}
	To determine the AGN nature of the galaxy companions, we use X-ray data from $Chandra$, $XRT$, or $XMM$.  Because of the difficulty identifying dual X-ray point sources at separations $<$20$\arcsec$ with $XMM$ or $XRT$, we obtained $Chandra$ observations (PI: Mushotzky, 12700910)  for 11 BAT AGN with close companions ($<$20$\arcsec$).  The $Chandra$ exposure times ensured  $>$10 photons for a source with 10$^{41}$ erg s$^{-1}$ and a column density of $5\times10^{23}$ cm$^{-2}$; objects with lower column densities will have more counts. 
	
	To fit X-ray spectra we assumed a fixed Galactic photoelectric absorption \citep{Kalberla:2005p11562}, a floating photoelectric absorption component, and a power law ($F\propto E^{-\Gamma+2})$.    For source counts we use a 1.5$\arcsec$ (3 pixel) radius aperture in $Chandra$, 6$\arcsec$ (1.5 pixel) radius in $XMM$, and 7.1$\arcsec$ (3 pixel) radius in $XRT$.  We apply aperture corrections based on the point spread function (PSF) at 1.5 keV. For sources with $<$50 counts, we use a fixed power law of $\Gamma=1.8$.  For sources without detections,  we calculate 95\% confidence limit Poisson statistics assuming $\Gamma=1.8$ and $N_{\mathrm{H}}=10^{22}$ cm$^{-2}$.  
	

\subsection{Emission Line Diagnostics and Optical Data}

	 We use optical spectroscopy from the SDSS DR8 to search for dual AGN within 100 kpc.  Because of the 55$\arcsec$ fiber collision limits in the SDSS, we observed 11 BAT AGN galaxies with Gemini.  We observed both galaxy nuclei simultaneously using the B600-G5307 grating with a 1$\arcsec$ slit in the 4300--7300 $\mathrm{\AA}$ wavelength range, for 37 minutes each. We follow \citet{Winter:2010p6825} for correcting Milky Way reddening, starlight continuum subtraction, and fitting AGN diagnostic lines.   To correct our line ratios for extinction, we use the narrow Balmer line ratio (H$\alpha$/H$\beta$) assuming an intrinsic ratio of 3.1 and the \citet{Cardelli:1989p1821} reddening curve.
	 
	 To compute the ratio of stellar masses of the AGN host galaxy and its companion, we use \textit{ugriz} photometry and follow \citet{Koss:2011p14514}.  If the galaxy nuclei were too close to accurately measure stellar masses in multiple filters, we use the ratio of $i$ band luminosity to determine stellar mass ratios or if unavailable, ratios of K band luminosity from 2MASS.

\begin{figure} 
\plotone{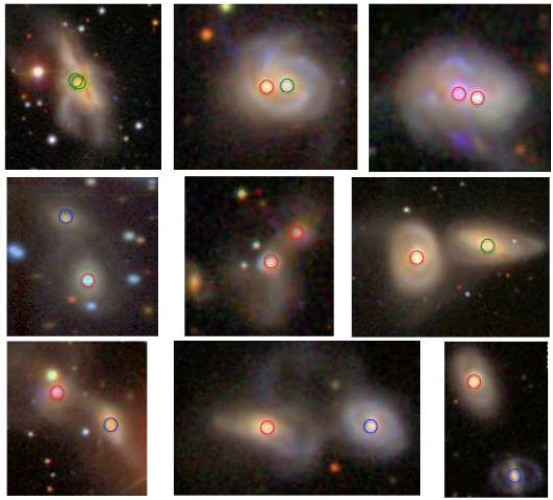} 
\caption{Nine composite \textit{gri} images selected at random from 16 BAT detected dual AGN in Table 1.  A red circle indicates X-ray and optical AGN detection.  A green circle indicates X-ray AGN detection, but no optical detection.  A blue circle indicates optical emission line diagnostics detect the AGN, but there are too few X-ray counts to detect an AGN.}
\label{dualpic}
\end{figure}

\subsection{Dual AGN Classification}
	We classify a source as a dual AGN when the companion galaxy's optical emission line diagnostics or X-ray data indicate an AGN.  We require that at least two of three emission line diagnostics indicate a Seyfert or LINER.  For the companions of BAT detected AGN, we also use X-ray data for companion classification.  Since X-rays can be produced from star formation, a galaxy is an AGN only if $L_{2-10 \: \mathrm{keV}}$$>$10$^{42}$ erg s$^{-1}$  or X-ray spectroscopy indicates a hard power law ($\Gamma$$<$$2$), an Fe K line, or rapid time variability indicative of an AGN.  At $L_{2-10 \: \mathrm{keV}}$$>$$10^{42}$ erg s$^{-1}$, a very high star formation rate (SFR$>$200 $M_\odot$~yr$^{-1}$) would be required to generate the observed hard X-ray luminosity \citep{Ranalli:2003p11744}.  For AGN with multiple companions within 100 kpc, we test each galaxy pair separately and only compare the three nearest companions to ensure the sample is not dominated by galaxy groups.  We note this AGN classification excludes some lower luminosity AGN with lower X-ray luminosities or composite optical spectra.
\section{Results}

\subsection{Companion Sample Study}
	We identify 81/167 (49\%) of BAT AGN having at least one companion within 100 kpc.   Additionally, 16 BAT AGN galaxies have two companions within 100 kpc and 5 have three or more companions within 100 kpc.  The total number of companions in the sample is 106.  A list of the BAT AGN and their companions can be found in Table 1. In the SDSS Seyfert sample, 358/1988 (18\%) have at least one companion within 100 kpc.  Additionally, 17 SDSS Seyferts have two companions within 100 kpc and 5 have three or more within 100 kpc, for a total of 385 companions. 
	
	We find that major mergers ($M_1/M_2$$<$3) are more frequent at close separations for BAT and SDSS Seyferts.  At $<$15 kpc, 78\% (14/18)  of BAT AGN galaxy companions have $M_1/M_2$$<$3, while at  60-100 kpc separations, only 30\% (10/33) have $M_1/M_2$$<$3.  Both the SDSS and BAT AGN sample show similar distributions of average stellar mass ratio with host galaxy separation. We also find that BAT AGN are more likely than SDSS Seyferts or inactive galaxies to have companions at small separations.  Namely, at separations $<$15 kpc, 10\% (16/167) of BAT AGN have companions, while for the SDSS Seyferts this fraction is only 1\% (20/1988), and only 2\% (4/167) among inactive galaxies matched in stellar mass and redshift to the BAT AGN consistent with \citet{Koss:2010p7366}.

\subsection{Dual AGN Frequency with Projected Separation}

	Among the entire sample of BAT AGN and SDSS Seyferts, we find a lower limit of the dual AGN frequency between 1 and 100 kpc of 9.6\% (16/167) for BAT AGN and 0.5\% (10/1988) for SDSS Seyferts.   These are lower limits because we exclude sources classified as composites by their optical emission line properties, and because both samples have incomplete X-ray and optical spectroscopy. 
	
	Sample images of the BAT detected dual AGN can be found in Figure~\ref{dualpic}.  The majority of these dual AGN (75\%, 12/16) are found on scales $<$30 kpc.  NGC 835 is a triple AGN system with AGN companions at 15 and 87 kpc.  A plot of the frequency of dual AGN by separation is in Figure \ref{dualfrac}.  The dual AGN fraction peaks at small separations ($<$15 kpc), where 50\% (8/16) are duals.  

	Dual AGN in the BAT and SDSS have, on average, smaller separations than the AGN with inactive companions.  The mean and 1$\sigma$ values of the separations among BAT AGN and their companions are 28$\pm$24 kpc, and 50$\pm$26 kpc for the dual AGN and the AGN-inactive galaxy systems, respectively.  A Kolmogorov-Smirnov (K-S) test indicates a negligible ($<$2$\times$10$^{-6}$) chance that the distribution of separations for the dual AGN and AGN-inactive systems are from the same distribution. A similar result is found for SDSS Seyferts with separations of 41$\pm$30 kpc, and 64$\pm$24 kpc for the dual AGN and single AGN systems, respectively, with a $<$5\% chance from a K-S test.  The BAT AGN in duals are at smaller separations than SDSS Seyferts in duals with a $<$0.01\% chance of being from the same distribution.

\subsection{Host Properties of Dual AGN}

\begin{figure} 
\includegraphics[width=7.0cm]{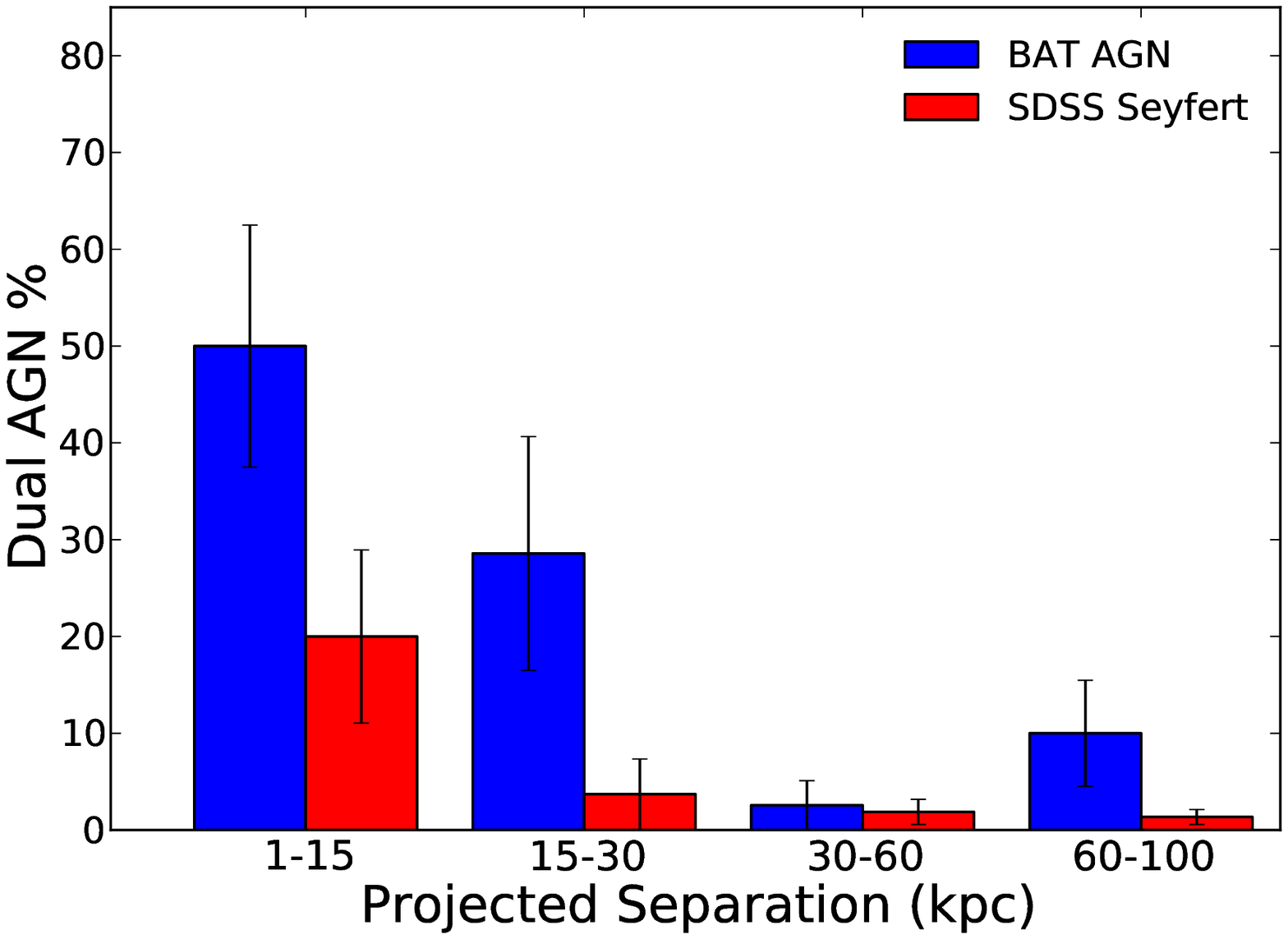}
\includegraphics[width=7.0cm]{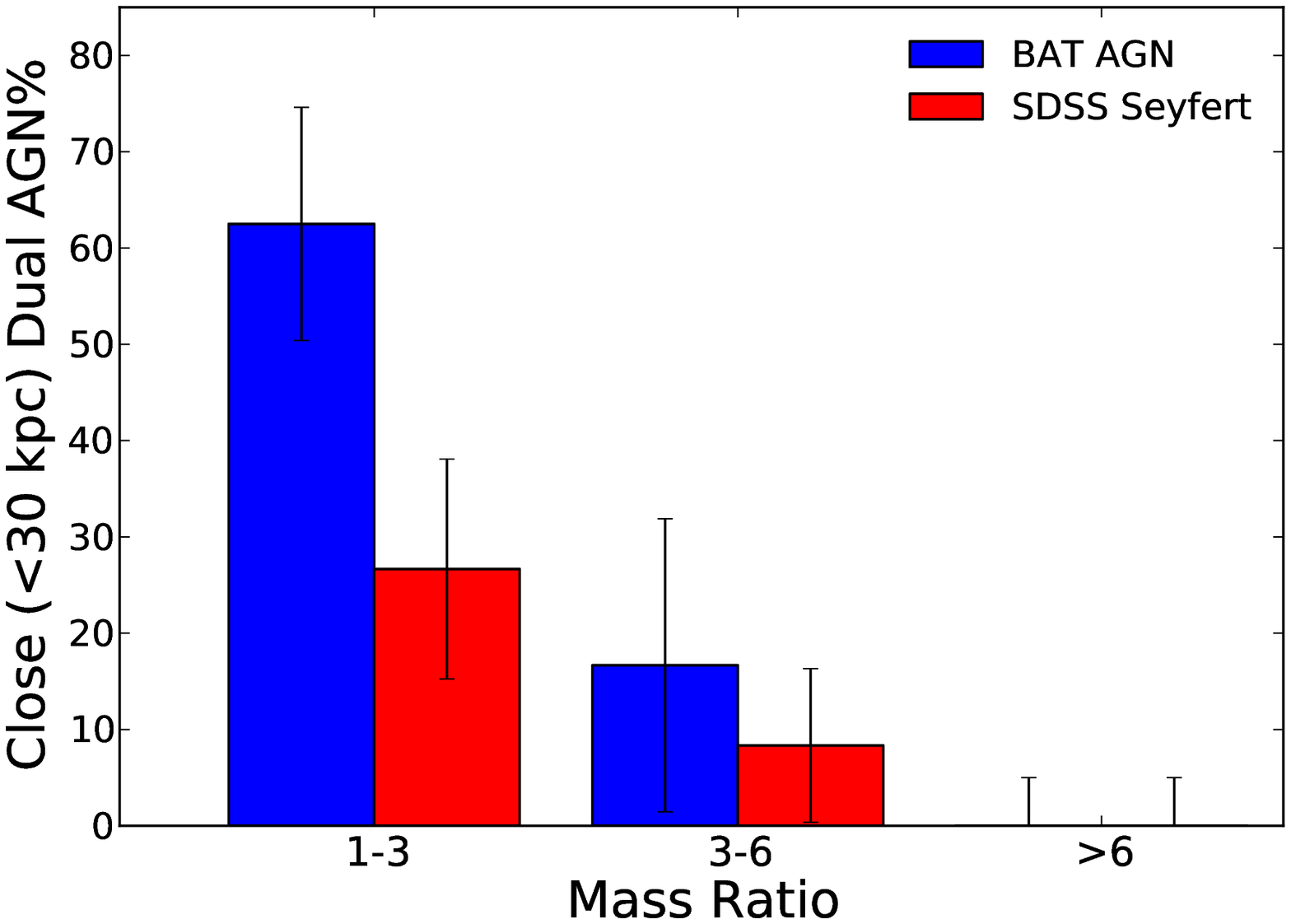} 
\caption{{\em Top:}  frequency of dual AGN for galaxies with companions as a function of apparent separation. Error bars assume binomial statistics.  We find the dual AGN frequency increases at smaller separations in BAT AGN and SDSS Seyferts.  {\em Bottom:} frequency of dual AGN as a function of the ratio of galaxy stellar masses.  We limited this study to projected separations $<$30 kpc because of the low frequency of duals found at larger separations.  The dual AGN frequency increases in major mergers for both BAT AGN and SDSS Seyferts.}
\label{dualfrac}
\end{figure}

	We studied the galaxy morphologies and merger types of systems hosting dual AGN.   We limited this study to projected separations $<$30 kpc because of the low frequency of duals found at larger separations and the higher chance of projection effects causing the galaxies to appear to be at artificially small separations.  We find that the frequency of dual AGN is strongly dependent on galaxy mass ratio (Figure \ref{dualfrac}) suggesting a minor merger is insufficient to trigger dual AGN.  The mean stellar mass ratio and 1$\sigma$ are  2.1$\pm$1.3.  The highest fraction of duals occurs in galaxy pairs with small stellar mass ratios ($M_1/M_2$$<$3) where 65$\%$ (11/17) of these pairs are in dual.  No dual AGN are found in galaxy pairs with $M_1/M_2$$>$$6$.

\begin{figure} 
\plotone{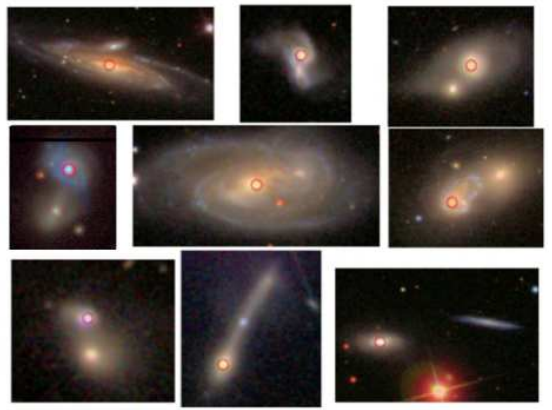} 
\caption{Composite \textit{gri} filter images of nine BAT AGN hosts with inactive companions within 30 kpc. A red circle indicates the BAT AGN. We find the majority of inactive companions are elliptical absorption line galaxies or minor mergers.}
\label{nonpic}
\end{figure}

	We also investigated the types of galaxies that are single AGN systems even though they have a close companion ($<$30 kpc) and a small stellar mass ratio ($<$6).  This includes a total of 11 systems (Figure \ref{nonpic}).  In optical spectroscopy, five of these companion galaxies are red elliptical absorption line galaxies, two are star forming emission line galaxies, and four have no available optical spectroscopy.   In the X-rays, 4/11 are at separations $<$20$\arcsec$, where a faint dual AGN may be missed and have no $Chandra$ observations.    The two star forming emission line galaxies not classified as duals are UGC 4727 and SDSS J112648.65+351454.2.  UGC 4727 is at stellar mass 5.9 times lower than the BAT source and a separation of 26 kpc.  UGC 4727 is also a bulgeless galaxy.  The other galaxy with star forming emission line diagnostics is SDSS J112648.65+351454.2 with a stellar mass 3.6 times lower than the BAT AGN, Mrk 423.   These results suggest that dual AGN tend to avoid elliptical/absorption line galaxies.     

\subsection{X-ray Properties of BAT AGN in Duals}

	We detect 12/16 of the dual AGN using X-rays.  Four of the secondary AGN are detected using optical spectroscopy, but have no X-ray detection.  One of these, UGC 11185, has a large upper limit because of its close proximity (28$\arcsec$) to a bright BAT AGN.  The remaining dual AGN with X-ray non-detections are at large separations ($>$40 kpc) and have a lower stellar mass than the BAT AGN (M$_2$/M$_1$$>$3).  The median value of the $L_{2-10 \: \mathrm{keV}}$ ratio between the dual AGN is 11, but varies dramatically, with the smallest ratio being 1.1 between NGC 6240S and NGC 6240N,  and some dual pairs having X-ray ratios greater than 1000 (IRAS 05589+2828 and UGC 8327).   
	
\begin{figure} 
\includegraphics[width=8.1cm]{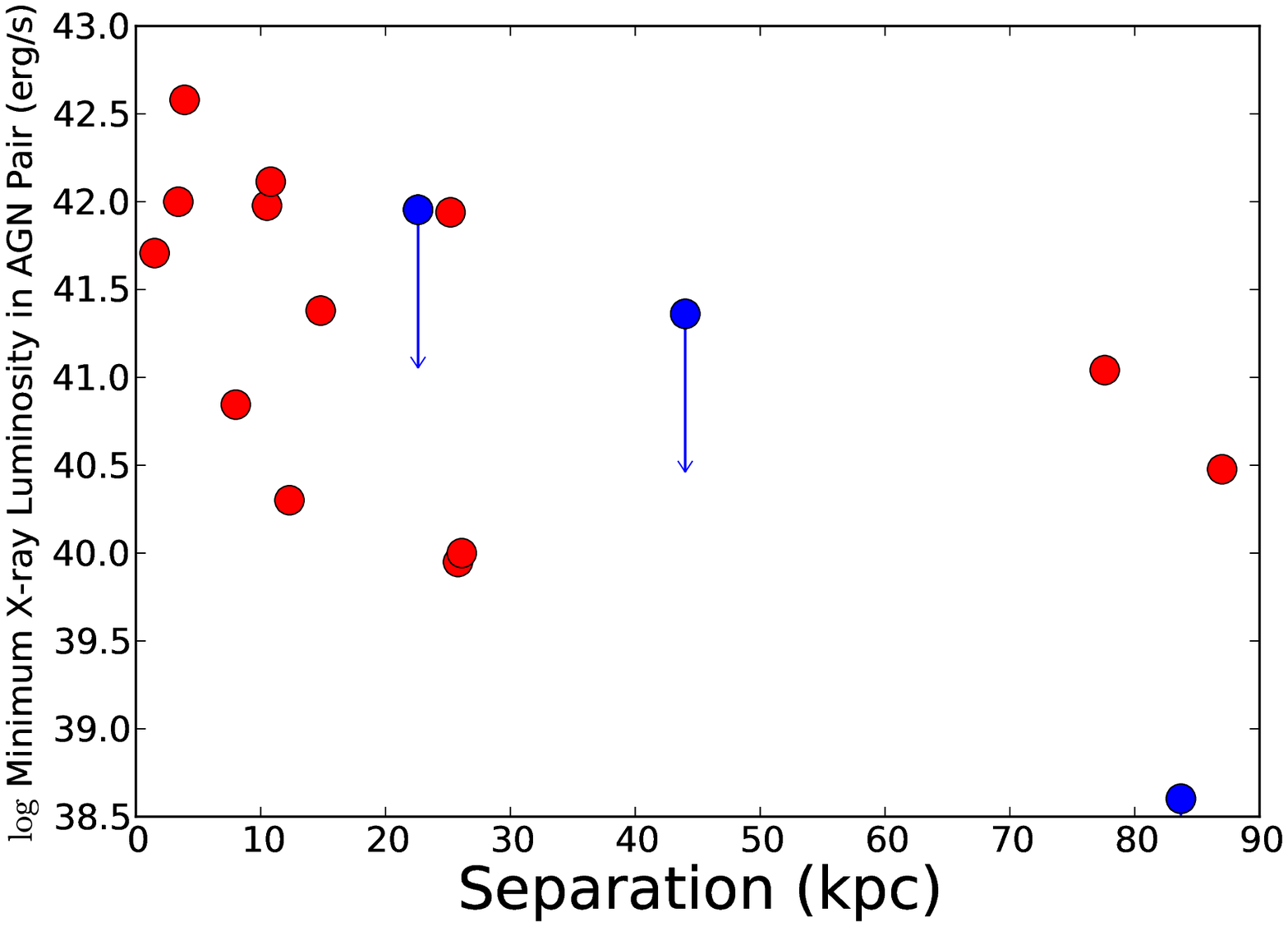}
\includegraphics[width=8.1cm]{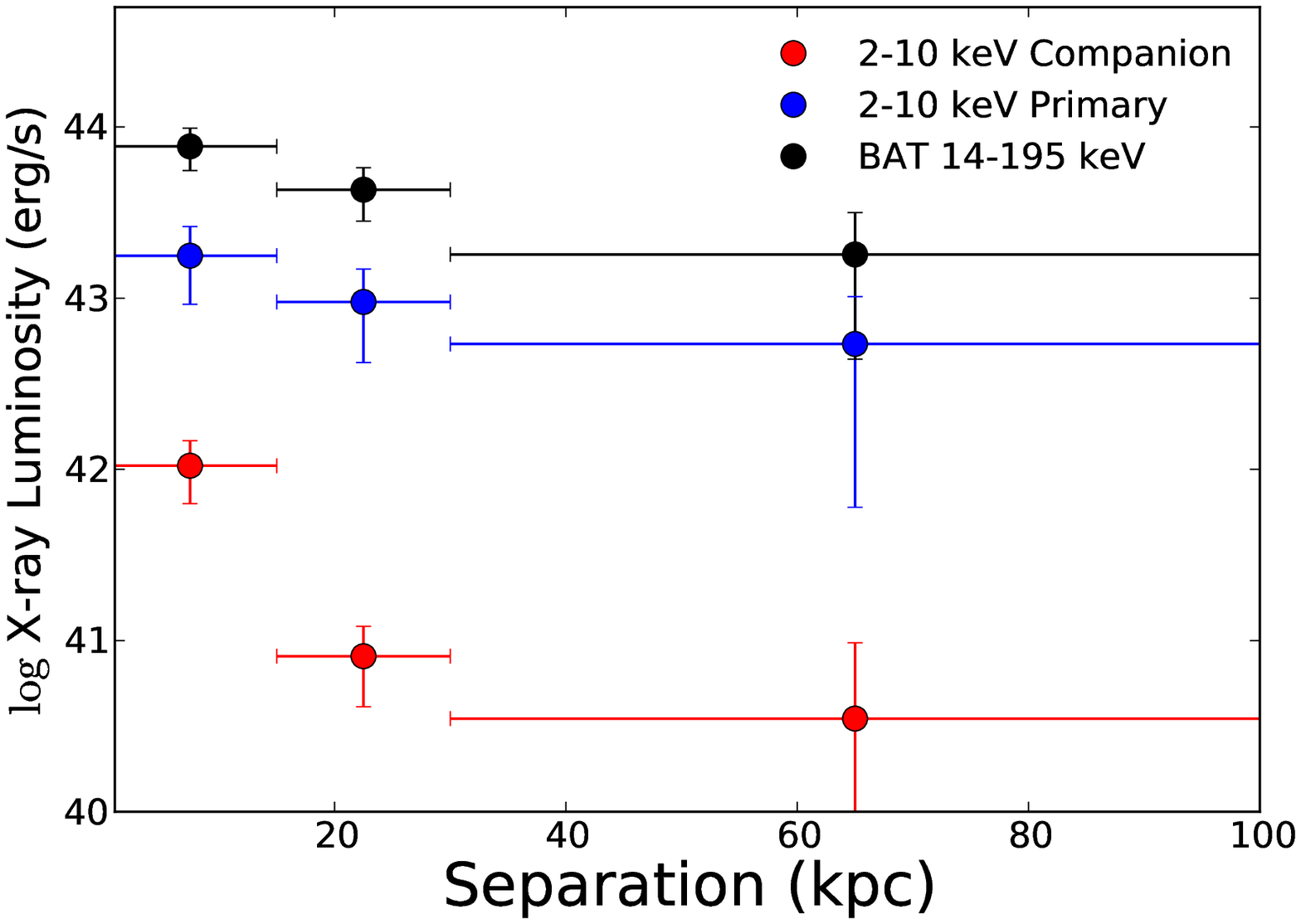}
\caption{{\em Top:} plot of minimum $L_{2-10 \: \mathrm{keV}}$ in 16 dual AGN pairs by galaxy separation.  Blue points are upper limits.  {\em Bottom:} binned average of X-ray luminosity with separation.  The X-ray luminosity of the primary and secondary AGN increases towards smaller separations.}
\label{dualxray}
\end{figure}

	We find that the X-ray luminosity of both AGN increases with decreasing separation, suggesting that the merging event is key to powering both AGN (Figure \ref{dualxray}).  For example, the six most luminous companion AGN are found at small separations ($<$15 kpc).  A Spearman correlation test indicates a $>$99.998\% probability that the $L_{2-10 \: \mathrm{keV}}$ of the AGN companion and separation are inversely correlated.  Similarly, a Spearman test indicates a $>$90\% that the primary AGN 2-10 keV luminosity and separation are inversely correlated. Finally, a Spearman test indicates a  $>$98.8\% chance the ultra hard X-ray luminosity is inversely correlated with separation, however, this low spatial resolution measure may be biased towards small separations because it includes emission from both AGN.
	
	
	Finally, we investigated the close ($<$30 kpc) dual AGN fraction with bolometric luminosity assuming a conversion factor of  $L_{\rm{bol}}=15\times L_{14-195\,\rm{keV}}$ based on \citet{Vasudevan:2010p5970}.  For systems with $L_{\rm{bol}}$$>$10$^{45}$ erg s$^{-1}$, the mean bolometric luminosity is $(2.2\pm1.1)\times$10$^{45}$ erg s$^{-1}$ and 17.1\% (6/35) are close ($<$30 kpc) dual AGN.  Among less luminous systems, the mean bolometric luminosity is $(4.4\pm2.9)\times$10$^{44}$ erg s$^{-1}$ and dual fraction is 4.5\% (6/132).  This suggests that close dual AGN may be more common among luminous systems.
	
\section{Summary and Discussion}	  
We study the frequency of dual AGN in nearby ultra hard X-ray selected BAT AGN using archival optical and X-ray imaging spectroscopy along with new observations from Gemini and $Chandra$.  We find:

\begin{enumerate}
\renewcommand{\theenumi}{(\roman{enumi})}
\renewcommand{\labelenumi}{\theenumi}
\item Dual AGN are much more likely to occur in systems with a close companion within $<$30 kpc among BAT AGN and SDSS Seyferts.  Among BAT AGN with a companion at small separations ($<$15 kpc), a high fraction (50\%, 8/16) are duals.   

\item In 19\% (3/16) of the dual AGN in the BAT sample, X-ray spectroscopy reveals the presence of a dual AGN that is not detected with emission line diagnostics (NGC 6240, Mrk 739, NGC 833).  

\item The hard X-ray luminosities of both AGN increase as the separation between the galaxies decrease suggesting that the merging event is key in powering the AGN.  Additionally, close dual AGN ($<$30 kpc) tend to be more common among the most X-ray luminous systems with $L_{\rm{bol}}$$>$10$^{45}$ erg s$^{-1}$.   This is in agreement with recent simulations suggesting the peak accretion and BH luminosity appear at small scales while relatively lower luminosity dual AGN are found at wider separations (Van Wassenhove et al.~2011). 

\item Dual AGN are more prominent among major mergers ($M_1/M_2$$<$3) and avoid absorption line galaxies with elliptical morphologies.  However,  AGN activity is more difficult to detect in dwarf companions because of the likely lower mass BHs.  

\item We find SDSS Seyferts are much less likely than BAT AGN (0.25\% vs.~7.8\%) to be found in dual AGN at scales $<$30 kpc.  

\end{enumerate}

	These results suggest that in systems already hosting a single AGN, nuclear activity in companions is triggered in the majority of close-separation major mergers.  {\em Indeed, our sample finds that all 11 BAT AGN with non-absorption line companions in major mergers ($M_1/M_2$$<$3) and close separations ($<$30~kpc) are dual AGN}.  Our sample also suggests that early-type gas-poor (red/elliptical) companions are unlikely to host AGN, possibly because it may not be able to capture enough gas during the merging process.  From these results, one might expect a high dual AGN fraction in luminous infrared galaxies (LIRGs) which are associated with gas-rich mergers. However, only a few dual AGN in this sample are LIRGs (e.g., NGC 6420, Mrk 463) which is consistent with the overall fraction of LIRGs in the rest of the BAT sample \citep[18\%,][]{Koss:2010p7366}.  Therefore, it seems that the particular type of merger triggering the luminous BAT AGN also triggers the secondary AGN. 
	
	
	
	
	Dual AGN systems are more common in ultra hard X-ray selected AGN than SDSS Seyferts since they have more close companions \citep{Koss:2010p7366}.   Since there are more dual AGN among BAT AGN than SDSS Seyferts by more than a factor of 10, the much higher fraction is most likely mainly because of the higher AGN luminosities of BAT AGN, not the lack of X-ray observations of SDSS Seyferts. 
	
	
	We also find a higher total fraction of dual AGN compared to dual AGN studies using the double peaked method \citep{Comerford:2009p15490}.  At scales $<$15 kpc, we find 5\% (8/167) are dual AGN, $\approx$10x higher than that estimated from double peak sources (0.3-0.5\%, at $z\approx0.4$, Rosario et al.~2011, $<3\%$, at $z\approx0.1-0.6$, Fu et al.~2011).  This is likely because:  1) double peak sources are at higher redshifts, $\approx$10 times that of this sample (z=0.02 vs.~z=0.3) so they are more difficult to detect as dual AGN, 2) the SDSS fiber is 3$\arcsec$, so it is sensitive to different, typically smaller physical scales (at $z=0.4$, 3$\arcsec$$\approx$15 kpc), 3) the double peak method is biased towards selecting certain types of single AGN such as radio bright AGN \citep{Smith:2010p7983}, which is different than BAT AGN where only 0.5\% are radio loud \citep{Teng:2011p15419}, 4) the double peak method has difficulty detecting dual AGN with small velocity offsets, weak lines, or those dual AGN that are detected only in the X-rays like NGC 6240.  Finally, because of the fixed SDSS fiber size, the double peak method is sensitive to AGN with different physical separations and different luminosities depending on the redshift range studied making a comparison difficult, but our results suggest that the fraction of dual AGN is much higher than the double peak method suggests among luminous AGN.  
	
	
	This type of extensive study is impossible to do at higher redshifts because of surface brightness dimming and the limitations in resolution and exposure time, particularly in X-rays since the secondary AGN are typically 10-100x fainter.  However, the pair fraction and frequency of gas-rich "wet" galaxy mergers increases with redshift.  This suggests that dual AGN activation may be much more common at higher redshifts.

\scriptsize
\begin{center}
\begin{longtable}{l l l l l l l l} 
\caption{\textbf{Dual AGN Study of BAT Sample}}  \label{f33} \\
\hline \hline \\[-2ex]
   \multicolumn{1}{c}{\textbf{BAT Source Name}} &
   \multicolumn{1}{c}{\textbf{Galaxy Comp.\tablenotemark{1}}} &
   \multicolumn{1}{c}{\textbf{Sep}} &
   \multicolumn{1}{c}{\textbf{$M_1/M_2$\tablenotemark{2}}} &
   \multicolumn{1}{c}{\textbf{X-ray 2\tablenotemark{3}}} &
   \multicolumn{1}{c}{\textbf{Opt 2\tablenotemark{4}  }} &
   \multicolumn{1}{c}{\textbf{Refs\tablenotemark{5}}} &
   \multicolumn{1}{c}{\textbf{Refs\tablenotemark{6}} } \\
   
   \multicolumn{1}{c}{} &
   \multicolumn{1}{c}{} &
   \multicolumn{1}{c}{\textbf{kpc}} &
   \multicolumn{1}{c}{} &
   \multicolumn{1}{c}{\textbf{2-10 keV}} &
   \multicolumn{1}{c}{\textbf{Diag}} &
   \multicolumn{1}{c}{\textbf{X-ray}} &
   \multicolumn{1}{c}{\textbf{Opt} }

\\[0.5ex] \hline
   \\[-1.8ex]
\endfirsthead

\multicolumn{3}{c}{{\tablename} \thetable{} -- Continued} \\[0.5ex]
   \multicolumn{1}{c}{\textbf{BAT Source Name}} &
   \multicolumn{1}{c}{\textbf{Galaxy Comp.\tablenotemark{1}}} &
   \multicolumn{1}{c}{\textbf{Sep}} &
   \multicolumn{1}{c}{\textbf{$M_1/M_2$\tablenotemark{2}}} &
   \multicolumn{1}{c}{\textbf{X-ray 2\tablenotemark{3}}} &
   \multicolumn{1}{c}{\textbf{Opt 2\tablenotemark{4}  }} &
   \multicolumn{1}{c}{\textbf{Refs\tablenotemark{5}}} &
   \multicolumn{1}{c}{\textbf{Refs\tablenotemark{6}}} \\
   
   \multicolumn{1}{c}{} &
   \multicolumn{1}{c}{} &
   \multicolumn{1}{c}{\textbf{kpc}} &
   \multicolumn{1}{c}{} &
   \multicolumn{1}{c}{\textbf{2-10 keV}} &
   \multicolumn{1}{c}{\textbf{Diag}} &
   \multicolumn{1}{c}{\textbf{X-ray}} &
   \multicolumn{1}{c}{\textbf{Opt} }
    \\[0.5ex] \hline
   \\[-1.8ex]
\endhead

\multicolumn{3}{l}{{Continued on Next Page\ldots}} \\
\endfoot
  \\[-1.8ex] \hline \hline
\endlastfoot

{\em Dual AGN Systems}\\
\hline
NGC 6240S& NGC 6240N& 1.5& 1.6& 0.51& ? \tablenotemark{7}& CXO& \\
Mrk 739E& Mrk 739W& 3.4& 0.5& 1& C/SF/SF& CXO& G\\
Mrk 463E& Mrk 463W& 3.9& 1.2& 3.8& AGN/Sy2/Sy2& CXO& S\\
IRAS 05589+2828& 2MASX J06021107+2828382& 8& 5.7& 0.07& AGN/Sy2/Sy2& CXO& G\\
ESO 509-IG 066 NED 02& ESO 509-IG066E& 10.5& 1.6& 0.95& AGN/Sy2/Sy2& XMM& 6DF\\
IRAS 03219+4031& 2MASX J03251221+4042021& 10.8& 2.6& 1.3& AGN/Sy2/Sy2& XMM& 4\\
NGC 3227& NGC 3226& 12.3& 1.7& 0.02& AGN/L/L& CXO& 1\\
NGC 835& NGC 833& 14.8& 1.5& 0.7& SF/L/SF& XMM& S\\
UGC 11185 NED02& UGC 11185 NED01& 22.6& 1.3& $<0.9$& AGN/Sy2/Sy2& XRT& G\\
MCG +04-48-002& NGC 6921& 25.2& 0.4& 0.87& Sy2& XMM& N\\
NGC 2992& NGC 2993& 25.8& 2.7& 0.0089& SF/SF/SF& XMM& 6DF\\
UGC 8327 NED02& UGC 8327 NED01& 26.1& 2.8& 0.014& AGN/Sy2/Sy2& CXO& S\\
Mrk 268& MRK 268SE & 44& 5.5& $<0.23$& AGN/Sy2/Sy2& XMM& S\\
NGC 7679& NGC 7682& 77.6& 2.3& 0.11& Sy2& XMM& 5\\
NGC 1052& NGC 1042& 83.7& 3.3& $<0.0004$& C/SF/Sy2& CXO& S\\
M106& NGC 4220& 85.7& 12& ?& AGN/Sy2/Sy2& & 1\\
NGC 835& NGC 839& 87& 2.1& 0.03& AGN/Sy2/?& XMM& 6\\
\hline
{\em Single AGN Systems}\\
\hline
Mrk 423& SDSS J112648.65+351454.2& 5.9& 3.6& $<1.8$& SF/SF/SF& XRT& S\\
NGC 931& LEDA 212995& 6& 6.1& $<0.0076$& HII& CXO& 3\\
ARP 151& SDSS J112535.23+542314.3& 8.2& 1.4& $<0.33$& Abs& XRT& S\\
2MASX J09043699+5536025& 2MASX J09043675+5535515& 9& 0.9& $<9.3$& Abs& XMM& G\\
NGC 235A& NGC 235B& 9& 5.5& $<0.011$& Abs& CXO& G\\
UGC 03995& UGC 03995A& 10.3& 8.2& $<0.0086$& SF/SF/SF& CXO& S\\
2MASX J11454045-1827149& 2MASX J11454080-1827359& 14& 1.2& $0.04$& Abs& CXO& G\\
NGC 3786& NGC 3788& 14& 1.2& $<0.015$& ?& XMM& \\
FAIRALL 0272& & 19& 15.6& $<0.038$& ?& XMM& G\\
Mrk 348& 2MASX J00485285+3157309& 22& 5.2& $0.0091$& ?& XMM& \\
KUG 1208+386& 2MASX J12104784+3820393& 24.1& 11& $<0.17$& SF/SF/SF& XMM& S\\
NGC 5106& PGC 046603& 25& 6.6& $<0.046$& SF/SF/SF& XRT& S\\
NGC 7469& IC 5283& 25& 4.5& $<0.0017$& ?& CXO& \\
UGC 05881& SDSS J104644.87+255502.1& 25.1& 20.9& $<0.0072$& SF/SF/SF& XRT& S\\
Mrk 18& UGC 04727& 25.9& 5.9& $<0.037$& SF/SF/SF& XMM& S\\
NGC 1142& NGC 1143& 26& 1.7& $<0.045$& Abs& XMM& S\\
Mrk 279& MCG +12-13-024& 27& 7.5& $<0.2$& ?& XMM& G\\
NGC 5506& NGC 5507& 27.2& 2.3& $<0.00085$& Abs& XMM& 6DF\\
UGC 07064& SDSS J120445.20+311132.9& 30.3& 5.2& $<0.019$& Abs& XMM& S\\
MCG +06-24-008& SDSS J104444.22+381032.9& 30.7& 33.9& $<0.077$& LowSN& XRT& S\\
NGC 3079& CGCG 265-055& 30.9& 47.9& $0.0016$& SF/SF/SF& XMM& S\\
UGC 07064& 2MASX J12044518+3111327& 31& 6.7& $<0.014$& ?& XMM& \\
MCG -01-24-012& MCG -01-24-011& 32.5& 0.9& $<0.18$& Abs& XMM& 6DF\\
MCG -02-12-050& 2MASX J04381113-1047474& 33& 5.1& $<0.4$& ?& XRT& \\
Mrk 590& SDSS J021429.36-004604.7& 33.8& 34.7& $0.007$& Abs& XMM& S\\
UGC 06527 NED03& UGC 06527 NOTES02& 35.7& 0.5& $<0.022$& SF/SF/SF& XMM& S\\
FAIRALL 272& Fairall 0273& 36& 4.8& $0.026$& ?& XMM& \\
NGC 5231& SDSS J133542.77+030006.7& 36.3& 25.7& $<0.017$& SF/SF/SF& XMM& S\\
MCG -01-13-025& 2MASX J04514177-0350295& 36.8& 5& $<0.050$& abs& XMM& 6DF\\
UGC 07064& SDSS J120445.27+310927.8& 36.9& 5.5& $0.051$& SF/SF/SF& XMM& S\\
LEDA 214543& NGC 6230 NED02& 38.5& 0.6& $<0.053$& ?& XRT& \\
Mrk 477& PGC 052445& 39& 1.6& $?$& ?& & \\
2MASX J09043699+5536025& 2MASX J09043156+5535326& 41& 0.9& $<0.64$& ?& XMM& \\
NGC 7319& NGC 7318A& 41& 0.7& $0.019$& ?& CXO& \\
NGC 4593& MCG -01-32-033& 41.1& 58.9& $<0.00097$& SF/SF/?& XMM& 6DF\\
UGC 06527 NED03& UGC 06527 NOTES01& 41.8& 2.6& $<0.011$& SF/SF/SF& XMM& S\\
Mrk 268& 2MASX J13411364+3023281& 44.2& 27.5& $<0.071$& SF/SF/SF& XMM& S\\
NGC 1194& 2MASX J03034116-0104249& 44.2& 4.2& $<0.0037$& SF/SF/SF& XMM& S\\
NGC 513& KUG 0121+335& 44.6& 9.3& $<0.033$& ?& XMM& \\
NGC 3079& NGC 3073& 45& 25.7& $<0.0021$& SF/SF/SF& CXO& 1\\
NGC 2992& FGC 0938& 45.5& 90.8& $<0.013$& ?& XMM& \\
SBS 0915+556& 2MASX J09191393+5528422& 47& 0.7& $<0.2$& SF/SF/SF& XRT& S\\
NGC 3718& NGC 3729& 47.7& 2.4& $0.0025$& ?& CXO& \\
2MASX J09112999+4528060& SDSS J091122.60+452717.2& 49& 9.1& $<0.0063$& SF/SF/SF& XMM& S\\
NGC 4388& VPC 0415& 49.2& 134.9& $<0.017$& LowSN& XMM& S\\
NGC 5290& SDSS J134529.67+413831.0& 49.4& 134.9& $<0.004$& LowSN& XRT& S\\
NGC 2885& UGC 05037 NOTES02& 52.9& 4& $<0.051$& Abs& XRT& S\\
NGC 235A& NGC 0232& 53.8& 0.8& $0.11$& AGN/SF/SF& CXO& 6DF\\
NGC 3081& TOLOLO 0957-225& 54& ?& $<0.0025$& ?& XRT& \\
NGC 1052& 2MASX J02413514-0810243& 54.8& 56.2& $<0.00036$& Abs& CXO& S\\
NGC 5252& SDSS J133811.34+043052.4& 54.8& 61.7& $<0.036$& SF& XMM& S\\
NGC 2885& MCG +04-22-060& 56.2& 3.3& $<0.052$& Abs& XRT& S\\
NGC 835& NGC 838& 56.7& 2.3& $<0.05$& SF/SF/SF& XMM& S\\
Mrk 926& 2MASX J23044397-0842114& 58.7& 3.3& $<0.56$& Abs& XMM& S\\
MCG -02-12-050& 2MASX J04381880-1047004& 58.9& 9.4& $<0.15$& ?& XRT& \\
Mrk 915& MCG -02-57-024& 59.6& 11.8& $<0.02$& ?& XRT& \\
SBS 1301+540& MCG +09-21-095& 60& 6.2& $<0.035$& SF/SF/SF& XMM& S\\
NGC 5290& SDSS J134548.55+414443.5& 61.7& 107.2& $<0.018$& LowSN& XRT& S\\
NGC 1052& NGC 1047& 62.4& 22.4& $<0.00055$& Abs& XMM& S\\
MCG -01-24-012& MCG -01-24-013& 63.6& 2.1& $<0.2$& SF/SF/SF& XMM& 6DF\\
NGC 1142& 2MASX J02550661-0009448& 66.6& 7.5& $<0.045$& Abs& XMM& S\\
KUG 1141+371& SDSS J114435.26+365408.8& 66.9& 12.3& $<0.14$& SF/SF/SF& XMM& S\\
M106& NGC 4217& 67& 6.3& $<0.0001$& Comp/SF/SF& CXO& 1\\
Mrk 464& SDSS J135550.25+383332.0& 67.7& 2.1& $<0.034$& SF/SF/SF& XMM& S\\
2MASX J07595347+2323241& SDSS J075959.87+232448.2& 69& 23.4& $<0.27$& SF/SF/SF& XMM& S\\
Mrk 1469& SDSS J121617.69+505020.4& 69.2& 13.2& $<0.063$& SF/SF/SF& XRT& S\\
Mrk 477& 2MASX J14403849+5328414& 71.2& 5& $?$& SF/SF/SF& & S\\
IC 0486& SDSS J080013.23+263525.2& 72.8& 32.4& $<0.050$& SF/SF/SF& XRT& S\\
NGC 3079& SDSS J100331.69+553121.1& 75.7& 151.4& $?$& SF/SF/SF& & S\\
UM 614& 2dFGRS N404Z023& 77.7& ?& $<0.61$& SF/Sy/?& XMM& 2DF\\
NGC 4388& VCC 0871& 79.5& 42.7& $<0.19$& LowSN& XMM& S\\
NGC 7319& NGC 7317& 80.9& 1.6& $0.011$& ?& CXO& \\
2MASX J11454045-1827149& 1H 1142-178:[KPC2006] 2& 82.9& ?& $<0.038$& ?& CXO& \\
Mrk 817& 2MASX J14362084+5845279& 83.1& 4.1& $<0.34$& SF/SF/SF& XMM& S\\
MCG +02-21-013& SDSS J080440.36+104513.0& 83.4& 47.9& $<0.07$& SF/SF/LowSN& XRT& S\\
Mrk 352& CGCG 501-056& 85.6& 2.1& $<0.008$& ?& XMM& \\
NGC 973& IC 1815& 88.6& 2.2& $<0.034$& ?& XRT& \\
Mrk 1392& CGCG 048-116& 88.7& 1& $<0.31$& ?& XRT& \\
NGC 4686& SDSS J124609.71+543144.8& 89.6& 136& $<0.0048$& SF/SF/SF& XMM& S\\
ARK 347& SDSS J120425.69+201548.9& 89.9& 6.3& $<0.031$& Abs& XRT& S\\
CGCG 319-007& MCG +11-19-005& 94.5& 2.2& $<0.1$& ?& XRT& \\
Mrk 704& SDSS J091832.18+161604.0& 94.8& 55& $<0.07$& SF/SF/SF& XRT& S\\
NGC 4619& SDSS J124134.51+350634.6& 97.4& 100& $?$& SF/SF/SF& & S\\
NGC 5899& NGC 5900& 98.7& 2.2& $<0.0004$& ?& XMM& \\
UGC 03995& SDSS J074349.16+291154.5& 99& 36.9& $<0.01$& Abs& CXO& S\\
NGC 973& 2MFGC 02027& 99.1& 19.3& $?$& ?& & 
\footnotetext[1]{Projected separation.}
\footnotetext[2]{Host galaxy stellar mass ratio between BAT AGN and companion.}
\footnotetext[3]{Galaxy companion $L_{2-10 \: \mathrm{keV}}$ in units of 10$^{42}$ erg s$^{-1}$.}
\footnotetext[4]{AGN diagnostics from [O III] $\lambda$5007/H$\beta$ vs.~[N II] $\lambda$6583/H$\alpha$, [S II] $\lambda\lambda$6717, 6731/H$\alpha$, and [O I] $\lambda$6300/H$\alpha$ ratios.}
\footnotetext[5]{CXO=$Chandra$, XRT=$Swift$.}
\footnotetext[6]{Optical spectroscopy where S=SDSS, G=Gemini, N=NED, 1=\citet{Ho:1997p5224}, 2=\citet{Winter:2010p6825}, 3=\citet{Veilleux:1987p1782}, 4=Trippe in prep, 5=\citet{Veilleux:1995p14966}, 6=\citet{deCarvalho:1999p15478}}
\footnotetext[7]{? indicates no available X-ray or optical spectroscopy for the companion.}
\end{longtable}
\end{center}

\end{document}